\documentclass[12pt,a4paper]{article}
\usepackage[super]{cite}
\usepackage{graphics}
\usepackage{graphicx}% Include figure files
\usepackage{bm}% bold math
\usepackage{amsmath}
\usepackage{nccmath}
\usepackage[dvipdfmx]{hyperref}% add hypertext capabilities
\usepackage[dvipdfmx]{xcolor}
\usepackage{mathtools}
\usepackage{cite}
\usepackage{here}
\usepackage{color}
\usepackage{geometry} % 
\begin{document}

\markboth{S. Amano, Y. Aritomo \& M. Ohta}{Dynamics of neck formation in the coalescence of heavy nuclei}

%%%%%%%%%%%%%%%%%%%%% Publisher's Area please ignore %%%%%%%%%%%%%%
%\catchline{}{}{}{}{}
%%%%%%%%%%%%%%%%%%%%%%%%%%%%%%%%%%%%%%%%%%%%%%%%%%%%%%%%%%%%%%%%%%%

\title{Nuclear viscosity estimated by dynamics of neck formation in the early stage of nuclear collision
}

\author
{S. Amano$^{1}$, Y. Aritomo$^{1}$, M.Ohta$^{2}$\\
\scriptsize{$^{1}$Kindai University Higashi-Osaka, Osaka 577-8502, Japan}\\
\scriptsize{$^{2}$Konan University Kobe, Hyogo 658-8501, Japan}\\
\scriptsize{e-mail: amano.shota3@gmail.com}\\% Your email address
}

\date{}% Include the date command, but leave its argument blank.
\maketitle

%\author{Shota Amano\\
%\href{mailto:amano.shota3@gmail.com}{amano.shota3@gmail.com}
%}
%\address{Department of Science and Engineering, Kindai University, Higashi-Osaka, Osaka 577-8502, Japan
%}

%\author{Yoshihiro Aritomo}
%\address{Department of Energy and Materials, Kindai University, Higashi-Osaka, Osaka 577-8502, Japan
%}

%\author{Masahisa Ohta}
%\address{Department of Physics, Konan University, Kobe, Hyogo 658-8501, Japan
%}

%\maketitle

%\pub{Received (Day Month Year)}{Revised (Day Month Year)}

\begin{abstract}
The very early stage of the coalescence of two nuclei is studied and used to estimate the nuclear viscosity. The time evolution of the neck region has been simulated by the unified Langevin equation method, which is used in the analysis of heavy-ion collisions from the approaching stage to the fusion-fission stage. It is found that the transition from viscous to inertial coalescence that appeared in the neck growth of macroscopic drops can also be seen in the present simulation in nucleus--nucleus collisions. The dynamics of neck growth is analyzed in terms of the hydrodynamical formula and the viscosity coefficient of nuclear matter is estimated using the analogy of macroscopic drops.
%The validity of the liquid drop model in nuclear physics is also noted.

%\keywords{Langevin; Neck radius; viscosity coefficient; viscous-inertial crossover.}
\end{abstract}

%\ccode{25.70.-z, 47.55.nk}

\section{Introduction}\label{sec1}

The coalescence and rupture of liquid drops that we usually encounter in natural phenomena have been attracting considerable attention. Studies to elucidate these phenomena have a long history in the hydrodynamical field, starting from the 19th century \cite{Thomson86}. Since then, the knowledge on the coalescence and rupture of drops has been applied to industrial devices. The nature of liquid drops is also extensively investigated in the field of microscopic matters such as nuclei. In nuclear physics, the liquid drop model has been widely used to predict the nature of nuclei composed of nucleons. Some examples are Weizs\"{a}cker's mass formula \cite{Weizs35,Bethe36} and the transient state method by Bohr--Wheeler \cite{PhysRev.56.426} for the prediction of fission. These are part of the foundation of present-day nuclear physics.

In recent hydrodynamical studies, the mechanism of neck formation in the early stage of the coalescence of macroscopic drops has been elucidated extensively with the recent development of experimental technologies \cite{PhysRevLett.95.164503, PhysRevLett.98.224502,PhysRevLett.106.114501}. These studies have shown that in a short time duration after contact, viscous forces are expected to dominate the flow of the neck region, and the neck radius evolves in proportion to time $t$ from neck opening, but after that, in the inviscid limit, the neck radius increases in proportion to $t^{1/2}$ owing to the effect of the motion which is dominated by an inertial capillary force.
The crossover point between two regions is clearly identified and studied quantitatively \cite{PhysRevLett.98.224502,PhysRevLett.106.114501}.

The dynamical properties of a growing neck in nuclear collisions have been investigated in terms of the mass parameter for the neck degree of freedom \cite{Adam00}, and presented the difference of the neck formation by the microscopic mass parameter and the macroscopic Werner-Wheeler one \cite{Adam00,PhysRevC.13.2385}. We use here the macroscopic mass parameter and study the process of neck formation just after contact.

One of the aims of the present study is to investigate how neck formation proceeds in very early stage of the collision, and to confirm that the viscous-inertial transition appearing in macroscopic drop also occurs in microscopic drops such as nuclei.
%Therefore, our calculations in this study will contribute to enforcing the validity of the liquid drop concept in nuclei.
 At the same time, the viscosity of the nucleus is estimated from the evolution of neck formation.  Thus far, the viscosity of the nucleus has been determined from experimental data in the giant dipole resonance (GDR) \cite{PhysRevLett.69.249,BAUMANN1998428,PhysRevLett.82.3404,article,PhysRevLett.103.172501}, and its value is scattered around 0.025 TP (1 TP = $10^{12}$ P = $6.24\times10^{-22}$ MeV s/fm$^{3}$). P (poise) is the unit of  viscosity. In relation to fission phenomena such as the total kinetic energy of fission fragments, fusion cross sections, and pre-neutron emission multiplicity, various values of viscosity are required for fitting experiments \cite{PhysRevLett.70.3538}. In the analysis of fission processes, the viscosity is shown to be in the range of 0.04 TP$-$0.08 TP by comparing the microscopic and macroscopic dissipation energies \cite{KOONIN197787, PhysRevC.17.646}. The strong viscosity is inferred from the fission transition time \cite{PhysRevC.13.2385}.

In the recent theoretical analysis \cite{Zagrebaev_2005,PhysRevC.106.024610} of the deep inelastic collision (DIC), the quasifission (QF) and the fusion-fission (FF) processes, the viscosity or the friction force is treated in general as a model parameter. To clarify the uncertainty of the viscosity or the friction force, in connection with the analysis of QF processes, we investigated the nuclear viscosity in relation with the dynamics of neck formation in the contact stage of nucleus--nucleus collisions. We show that our results are consistent with the viscosity coefficient derived from GDR experiments.

In the following sections, a brief review of the method of our Langevin calculation used in this analysis formation in the early stage of collision is investigated in comparison with the case of macroscopic drops. The derivation of nuclear viscosity coefficient is also discussed.

%2023/1/11 amano add%%%%%%%%%%%%%
%In the present paper, the impact parameter $b$ is fixed 0 fm to bring the calculation conditions closer to the experimental conditions of drops coalescence.
%%%%%%%%%%%%%%%%%%%%%%%%

\section{Theoretical framework}\label{sec2}
\subsection{Potential energy}

Since the early stage of neck formation in nucleus--nucleus collisions cannot be seen directly like macroscopic drops can be, we use the simulation by the Langevin equation method \cite{Zagrebaev_2005,Zagrebaev_2007,ARITOMO20043} where the shape evolution of colliding nuclei is calculated in three--dimensional (3D) deformation space. In our Langevin method, the calculation begins with the separate configuration of two colliding nuclei. Then, two nuclei come into contact with each other to form a neck toward QF and FF. Therefore, for the potential energy in the 3D space, we consider the time evolution of the potential energy from the diabatic one $V_\text{diab}$ to the adiabatic one $V_\text{adiab}$\!.
The diabatic potential is calculated by a folding procedure using effective nucleon--nucleon interaction \cite{Zagrebaev_2005,Zagrebaev_2007,zagrebaev2007potential}.
However, the adiabatic potential energy of the system is calculated using an extended two-center shell model \cite{zagrebaev2007potential}.
%%%%% - check - %%%%%%%
Then, we connect the diabatic and adiabatic potential energies with a time-dependent weighting function $f\left(t\right)$ as
%%%%%%%%%%%%%%%%%
%
\begin{eqnarray}
&&V\left(q,t\right)=V_\text{diab}\left(q\right)f\left(t\right)+V_\text{adiab}\left(q,t,L,T\right)\left[1-f\left(t\right)\right], \\
&&f\left(t\right)=\exp{\left(-\frac{t}{\tau}\right)}.
\label{pot}
\end{eqnarray}
%
%%%%% - check - %%%%%%%
Here, $q$ denotes a set of collective coordinates representing a nuclear shape. $t$ is the interaction time and $\tau$ is the relaxation time in the transition from the diabatic potential energy to the adiabatic one. $L$ and $T$ denote the total angular momentum and the temperature of the compound nucleus calculated from the intrinsic energy of the composite system.
%%%%%%%%%%%%%%%%%
We use the relaxation time $\tau=0.1~\text{zs}$ proposed in Refs. \cite{bertsch1978collision,CASSING1983467,PhysRevC.69.021603}.
We use the two-center parameterizations as coordinates to represent nuclear deformation \cite{maruhn1972asymmetrie,sato1978microscopic}.
To solve the dynamical equation numerically and avoid the huge computation time,
we strictly limited the number of degrees of freedom and employed three parameters as follows:
$z_{0}=|z_{1}|+|z_{2}|$ (distance between the centers of two potentials),
$\delta$ (deformation of fragment), and $\alpha$ (mass asymmetry of colliding nuclei);
$\alpha=\frac{A_{2}-A_{1}}{A_\text{CN}}$,
where $A_{1}$ and $A_{2}$ not only stand for the mass numbers of the projectile and target, respectively \cite{Zagrebaev_2005,ARITOMO20043}, but also are then used to indicate the mass numbers of the two fission fragments. $A_\text{CN}$ is the mass number of the compound nucleus.
Figure 1 is the example of the nuclear shape described with the two-center parametrization.
The parameter $\delta$ is defined as $\delta=\frac{3\left(a-b\right)}{2a+b}$, where $a$ and $b$ represent the half  length of the ellipse axes in the $z_{0}$ and orthogonal to $z_{0}$ directions, respectively.
We assume that each fragment has the same deformation in the first step.
In addition, we use scaling to minimize the computation time and use the coordinate $z$ defined as $z=\frac{z_{0}}{R_\text{CN}B}$, where $R_\text{CN}$ denotes the radius of the spherical compound nucleus and the parameter $B$ is defined as $B=\frac{3+\delta}{3-2\delta}$.
The adiabatic potential energy is defined as
\begin{eqnarray}
%\begin{flalign}
&&V_\mathrm{{adiab}}\left(q,t,L,T\right)=V_\text{LDM}\left(q,t\right)+V_\text{SH}\left(q,T\right)+V_\text{rot}\left(q,L\right),
%V_\mathrm{{adiab}}\left(q,L,T\right)=V_\text{LDM}\left(q\right)+V_\text{SH}\left(q,T\right)+V_\text{rot}\left(q,L\right),
\label{adipot}
\end{eqnarray}
%\frac{\hbar^{2}\ell\left(\ell+1\right)}{2\mathcal{I}\left(q\right)} \nonumber \\
%+\frac{\hbar^{2}L_{1}(L_{1}+1)}{2\Im_{1}(q)} +\frac{\hbar^{2}L_{2}(L_{2}+1)}{2\Im_{2}(q)}, \nonumber \\
where $V_\text{LDM}$ and $V_\text{SH}$ are the potential energy of the finite-range liquid drop model and the microscopic energy that takes into account the temperature dependence, respectively.
%%%%%%%%%%%%%%%%%%%%%%%%
%
\begin{figure}[t]%fig1
\centering
\includegraphics[scale=0.40]{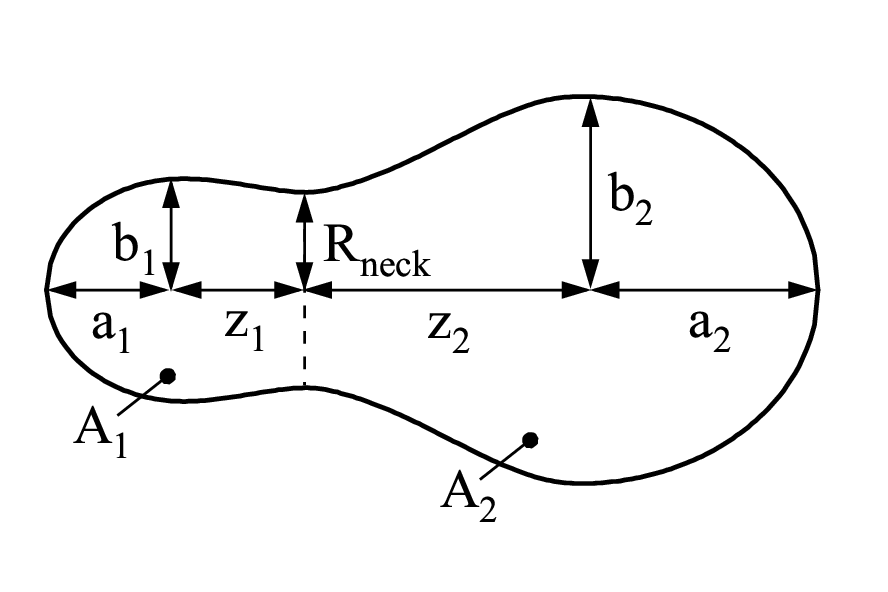}
\caption{Example of the nuclear shape described in the two-center shell model.
}\label{fig1}
\end{figure}
There is the neck parameter $\epsilon$ which is one of the parameter representing the nuclear shape.
The two harmonic oscillator potentials associated with the nuclear shape are smoothly connected by the neck parameter at the $z=0$ (See Fig. 2).
Toward the smaller value of $\epsilon$, the neck region is extended. The neck parameter is defined as
\begin{eqnarray}
\epsilon=E/E_{0}.
\label{epsilon}
\end{eqnarray}
%
%%%%%%%%%%%%%%%%%%%%%%%%

Concerning the adiabatic potential energy calculated using the two-center shell model, the neck parameter $\epsilon$ is included in the two-center parametrization. The value of the neck parameter $\epsilon$ is recommended as $0.35$ for fission processes \cite{YAMAJI1987487}. In our study, we treat $\epsilon = 1 $ as the entrance channel and $\epsilon = 0.35$ as the exit channel, and we assume the time dependence of $\epsilon$ to be,
\begin{eqnarray}
&&V_\text{LDM}\left(q,t\right)=V_\text{LDM}^{\epsilon=1}\left(q\right) f_{\epsilon}\left(t\right)+V_\text{LDM}^{\epsilon=0.35}\left(q\right) [1- f_{\epsilon}\left(t\right)], \\
&&f_{\epsilon}\left(t\right)=\frac{1}{1+\exp\left(\frac{t-t_{0}}{\Delta_{\epsilon}}\right)}.
\label{}
\end{eqnarray}
The temporal form of $\epsilon$ has been used commonly \cite{zagrebaev2007potential,sai22,sai19,kar17} and we adopted the characteristic relaxation time of the neck $t_0=0.35$ zs and the variance $\Delta_{\epsilon}=0.1$ zs in the present study.

%These values are determined considering the entrance channel characteristics : large Coulomb factor $Z_\text{p}Z_\text{t}$ and small degree of mass asymmetry of the colliding system. $Z_\text{p}$ and $Z_\text{t}$ denote the charges of the projectile and target, respectively.
%Details of the dynamical equation are described in our recent paper \cite{PhysRevC.106.024610}.
%
%
%-neck radius calculate%%%%%%%%%%%%%
%2023/1/11 amano add
In the two-center parametrization, the neck radius $R_\text{neck}$ is calculated from
\begin{eqnarray}
&&R_\mathrm{{neck}}=\sqrt{\frac{b_{1}^2-\left(\frac{z_{1}}{B}\right)^2 \epsilon}{1+G1}}\approx \sqrt{\frac{b_{1}^2-\left(\frac{z_{2}}{B}\right)^2 \epsilon}{1+G2}},  \\
%R_\mathrm{{neck}}&=&\sqrt{\frac{b_{1}^2-\left(\frac{z_{2}}{B}\right)^2 \epsilon}{1+G2}}, \nonumber \\
&&G1=\frac{z_{1}}{z_{2}-z_{1}}\left \lbrack1-\left(\frac{b_{2}}{b_{1}}\right)^2 \right \rbrack, ~
G2=-\frac{z_{2}}{z_{2}-z_{1}}\left \lbrack1-\left(\frac{b_{1}}{b_{2}}\right)^2 \right \rbrack.
\label{neckcal}
\end{eqnarray}
%
%%%%%%%%%%%%%%%%%%%%%%%%
\begin{figure}[t]%fig2
\centering
\includegraphics[scale=0.40]{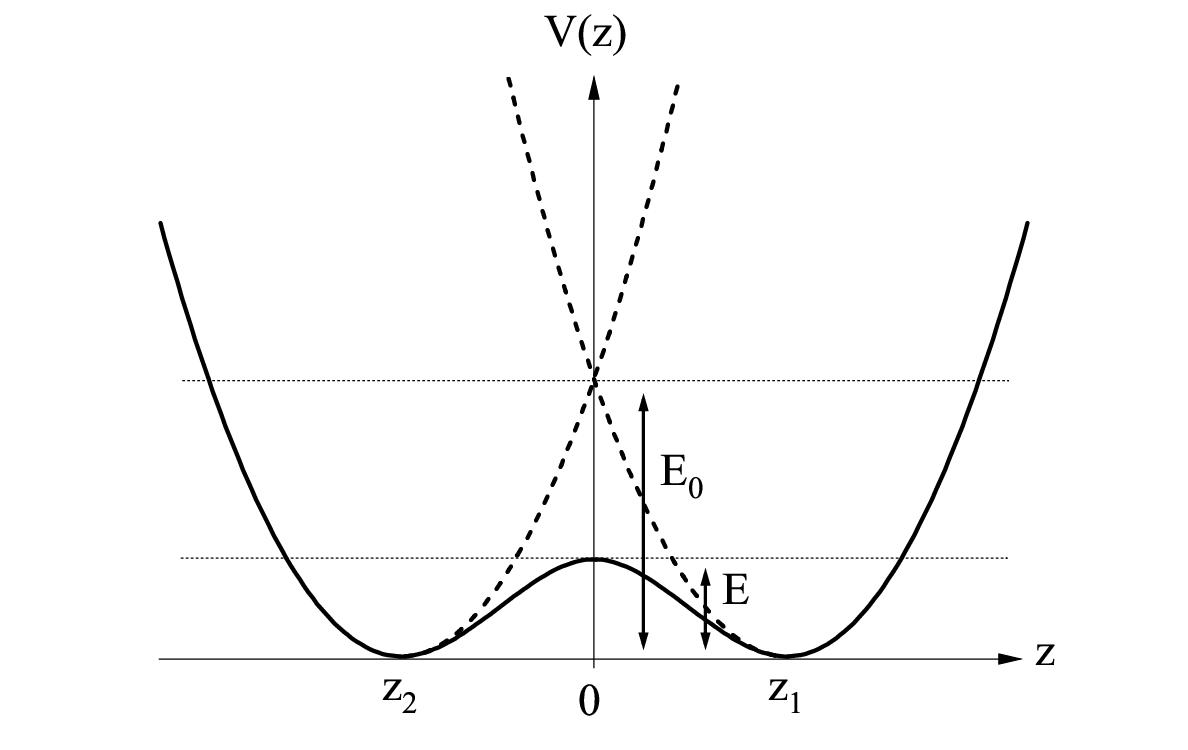}
\caption{Two harmonic oscillator potentials corresponding the nuclear shape. Two harmonic oscillator potentials are smoothly connected by the neck parameter $\epsilon=E/E_{0}$.
}\label{fig2}
\end{figure}
\begin{figure*}[t]%fig3
\centering
\includegraphics[scale=0.55]{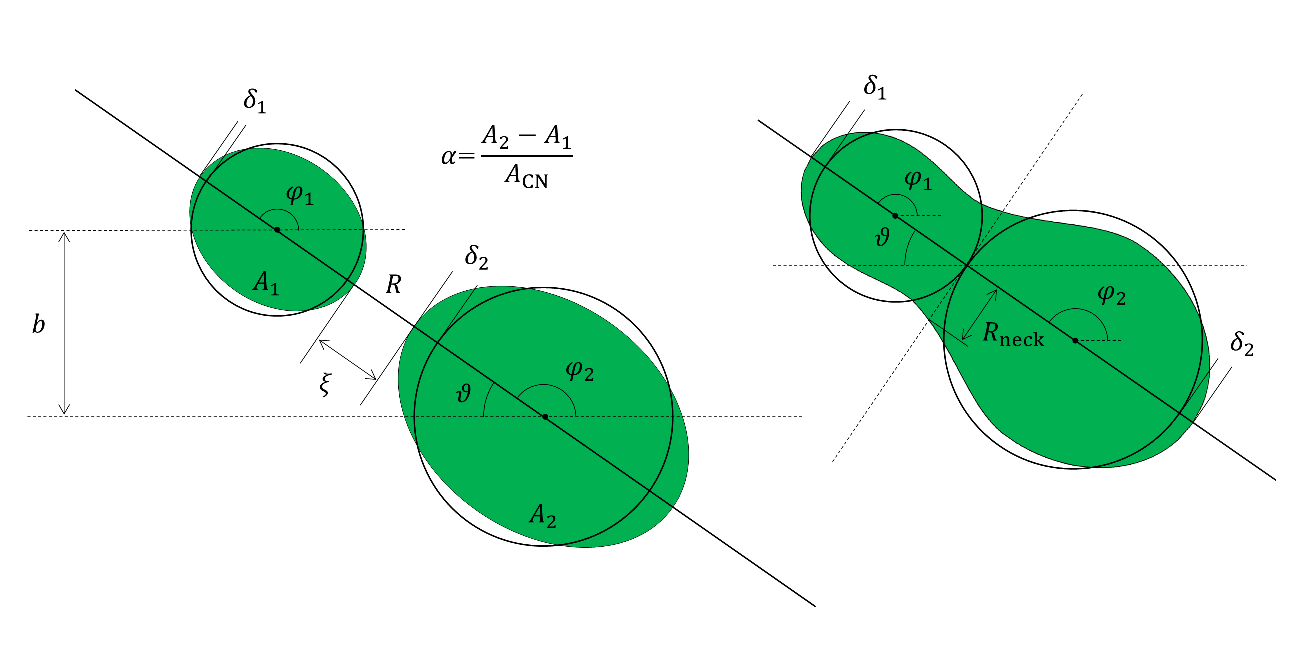}
\caption{Geometrical diagram with degrees of freedom used in the model.
}\label{fig3}
\end{figure*}
The simplified symbol $V_\text{LDM}$ and the symbol $V_\text{SH}$ are described as
\begin{eqnarray}
%\begin{fleqn}[2pt]
%\begin{align}
&&V_\text{LDM}\left(q\right)=E_\text{S}\left(q\right)+E_\text{C}\left(q\right), \\
&&V_\text{SH}\left(q,T\right)=E_\text{shell}^{0}\left(q\right)\Phi\left(T\right), \\
&&E_\text{shell}^{0}\left(q\right)=\Delta E_\text{shell}\left(q\right) + \Delta E_\text{pair}\left(q\right).
\end{eqnarray}
The symbols $E_\text{S}$ and $E_\text{C}$ stand for generalized surface energy \cite{PhysRevC.20.992} and Coulomb energy, respectively.
The symbol $E_\text{shell}^{0}$ indicates the microscopic energy at $T$ = 0, which is calculated as the sum of the shell
correction energy $\Delta E_\text{shell}$ and the pairing correlation correction
energy $\Delta E_\text{pair}$.
$\Delta E_\text{shell}$ is calculated by the Strutinsky method \cite{STRUTINSKY19681, RevModPhys.44.320}
from the single-particle levels of the two-center shell model
potential \cite{maruhn1972asymmetrie,suek74,10.1143/PTP.55.115} as the difference between the sum of
single-particle energies of occupied states and the averaged
quantity.
$\Delta E_\text{pair}$ is evaluated in the BCS approximation as described in Refs. \cite{RevModPhys.44.320, NILSSON19691}. The averaged part of the pairing correlation energy is calculated assuming that the density of single-particle
states is constant over the pairing window. The pairing strength constant is related to the average gap parameter $\tilde{\Delta}$ by solving the gap equation in the same approximation and adopting $\tilde{\Delta} = 12/ \sqrt{A}$ suggested in \cite{NILSSON19691} by considering the empirical results for the odd-even mass difference \cite{PhysRevC.90.054609}.
The temperature dependence factor $\Phi\left(T\right)=\exp\left(-\frac{E^{\ast}}{E_\text{d}}\right)$ is explained in Ref. \cite{ARITOMO20043}, where $E^{\ast}$ indicates the excitation energy of the compound nucleus. $E^{\ast}$ is given as $E^{\ast}=aT^{2}$, where $a$ is the level density parameter. The shell damping energy $E_\text{d}$ is selected as 20 MeV. This value is given by Ignatyuk et al. \cite{ignatyuk1975phenomenological}.
$V_\text{rot}$ is the centrifugal energy generated from the total angular momentum $L$. We obtain
\begin{eqnarray}
V_\text{rot}\left(q,L\right)=\frac{\hbar^{2}\ell\left(\ell+1\right)}{2\mathcal{I}\left(q\right)}+\frac{\hbar^{2}L_{1}(L_{1}+1)}{2\Im_{1}(q)}+\frac{\hbar^{2}L_{2}(L_{2}+1)}{2\Im_{2}(q)}.
%\Phi\left(T\right)=\exp\left(-\frac{E^{\ast}}{E_{d}}\right). \nonumber \\
%\end{align}
%\end{fleqn}
\end{eqnarray}
Here, $\mathcal{I}\left(q\right)$ and $\ell$ represent the moment of inertia of the rigid body with deformation $q$ and the relative orientation of nuclei and relative angular momentum respectively. The moment of inertia and the angular momentum  for the heavy and light fragments are $\Im_{1,2}$ and $L_{1,2}$, respectively.

%%%%%%%%%%%%%%%%%%%%
%                LANGEVIN
%%%%%%%%%%%%%%%%%%%%
\subsection{Langevin equation}
Langevin-type equations have seven degrees of freedom $\{R, \delta_{1}, \delta_{2}, \alpha, \vartheta, \varphi_{1}, \varphi_{2}\}$ as shown schematically in Fig. 3. However, in the present paper we set $\delta_{1}=\delta_{2}$.
We perform trajectory calculations of the time-dependent unified potential energy \cite{Zagrebaev_2005,Zagrebaev_2007,ARITOMO20043} using the multidimensional Langevin equation \cite{Zagrebaev_2005,ARITOMO20043,PhysRevC.80.064604} as follows :
%
%\begin{gather}
\begin{eqnarray}
&&\frac{dq_{i}}{dt}=\left(m^{-1}\right)_{ij}p_{j}, \nonumber \\
&&\frac{dp_{i}}{dt}=-\frac{\partial V}{\partial q_{i}}-\frac{1}{2}\frac{\partial}{\partial q_{i}}\left(m^{-1}\right)_{jk}p_{j}p_{k}-\gamma_{ij}\left(m^{-1}\right)_{jk}p_{k}+g_{ij}\Gamma_{j}\left(t\right), \nonumber \\
&&\frac{d\vartheta}{dt}=\frac{\ell}{\mu_{R}R^{2}}, \quad
\frac{d\varphi_{1}}{dt}=\frac{L_{1}}{\Im_{1}}, \quad
\frac{d\varphi_{2}}{dt}=\frac{L_{2}}{\Im_{2}}, \nonumber \\
&&\frac{d\ell}{dt}=-\frac{\partial V}{\partial\vartheta}-\gamma_\text{tan}\left(\frac{\ell}{\mu_{R}R^{2}}-\frac{L_{1}}{\Im_{1}}a_{1}-\frac{L_{2}}{\Im_{2}}a_{2}\right)R+R\sqrt{\gamma_\text{tan}T}\Gamma_\text{tan}\left(t\right), \nonumber \\
&&\frac{dL_{1}}{dt}=-\frac{\partial V}{\partial\varphi_{1}}+\gamma_\text{tan}\left(\frac{\ell}{\mu_{R}R^{2}}-\frac{L_{1}}{\Im_{1}}a_{1}-\frac{L_{2}}{\Im_{2}}a_{2}\right)a_{1}-a_{1}\sqrt{\gamma_\text{tan}T}\Gamma_\text{tan}\left(t\right), \nonumber \\
&&\frac{dL_{2}}{dt}=-\frac{\partial V}{\partial\varphi_{2}}+\gamma_\text{tan}\left(\frac{\ell}{\mu_{R}R^{2}}-\frac{L_{1}}{\Im_{1}}a_{1}-\frac{L_{2}}{\Im_{2}}a_{2}\right)a_{2}-a_{2}\sqrt{\gamma_\text{tan}T}\Gamma_\text{tan}\left(t\right).
%\label{lan}
%\end{gather}
\end{eqnarray}
The collective coordinates $q_{i}$ represent $z, \delta$, and $\alpha,$ the symbol $p_{i}$ denotes  momentum conjugated to $q_{i}$, and $V$ is the multidimensional potential energy. The symbol $\vartheta$ indicates the relative orientation of nuclei. $\varphi_{1}$ and $\varphi_{2}$ stand for the rotation angles of the nuclei in the reaction plane, $a_{1,2}=\frac{R}{2}\pm\frac{R_{1}-R_{2}}{2}$ is the distance from the center of the fragment to the middle point between the nuclear surfaces, and $R_{1,2}$ is the nuclear radii. The symbol $R$ is the distance between the nuclear centers.
The total angular momentum $L=\ell+L_{1}+L_{2}$ is preserved.

The symbol $\mu_{R}$ is reduced mass, and $\gamma_\text{tan}$ is the tangential friction force of the colliding nuclei.
Here, it is called sliding friction.
The phenomenological nuclear friction forces for separated nuclei are expressed in terms of $\gamma_\text{tan}$ and $\gamma_{R}$ for sliding friction and radial friction using the Woods-Saxon radial form factor described in Ref. \cite{Zagrebaev_2005}.
Sliding and radial friction are described as $\gamma_\text{tan}=\gamma_\text{t}^{0}F\left(\xi\right)$ and $\gamma_{R}=\gamma_{R}^{0}F\left(\xi\right)$, where the radial form factor $F\left(\xi\right)=\left(1+\exp^{\frac{\xi-\rho_{F}}{a_{F}}}\right)^{-1}$. $\gamma_\text{t}^{0}$ and $\gamma_{R}^{0}$ being the model parameters employed 0.1 $\times~10^{-22}$ MeV s fm$^{-2}$ and 100 $\times~10^{-22}$ MeV s fm$^{-2}$, respectively. $\rho_{F} \approx$ 2 fm and $a_{F} \approx$ 0.6 fm are also the model parameters determined in Ref. \cite{Zagrebaev_2005}, and $\xi$ is the distance between the nuclear surfaces $\xi=R-R_\text{contact}$, where $R_\text{contact}=R_{1}+R_{2}$ \cite{Zagrebaev_2005}.
The phenomenological friction for the radial direction is switched to the one-body friction in the mononucleus stage.
$\gamma_{R}$ is described to consider the kinetic dissipation according to the surface friction model \cite{FROBRICH1998131}
The radial friction is calculated as $\gamma_{zz}=\gamma_{zz}^\text{{one}}+\Omega\left(\xi\right)\gamma_{R}$.
For the mononuclear system, the wall-and-window one-body dissipation $\gamma_{zz}^\text{{one}}$ is adopted for the friction tensor \cite{PhysRevLett.70.3538,BLOCKI1978330,RAYFORDNIX1984161,RANDRUP1984105,Feldmeier_1987,CARJAN1986381,carj92,20067}.
$\Omega\left(\xi\right)$ is smoothing function switched the phenomenological friction to that of a mononuclear system as follows $\Omega\left(\xi\right)=\left(1+\exp^{-\frac{\xi}{0.3}}\right)^{-1}$ \cite{Zagrebaev_2005}.
$m_{ij}$ and $\gamma_{ij}$ stand for the shape-dependent collective inertia and friction tensors, respectively.
We adopted the hydrodynamical inertia tensor $m_{ij}$ in the Werner-Wheeler approximation for the velocity field \cite{PhysRevC.13.2385}.
The one-body friction tensors $\gamma_{ij}$ are evaluated within the wall-and-window formula \cite{RANDRUP1984105, PhysRevC.21.982}.
The normalized random force $\Gamma_{i}\left(t\right)$ is assumed to be white noise: $\langle \Gamma_{i} (t) \rangle$ = 0 and $\langle \Gamma_{i} (t_{1})\Gamma_{j} (t_{2})\rangle = 2 \delta_{ij}\delta (t_{1}-t_{2})$.
According to the Einstein relation, the strength of the random force $g_{ij}$ is given as $\gamma_{ij}T=\sum_{k}{g_{ik}g_{jk}}$.

\begin{figure}[t]%fig4
\centering
\includegraphics[scale=0.54]{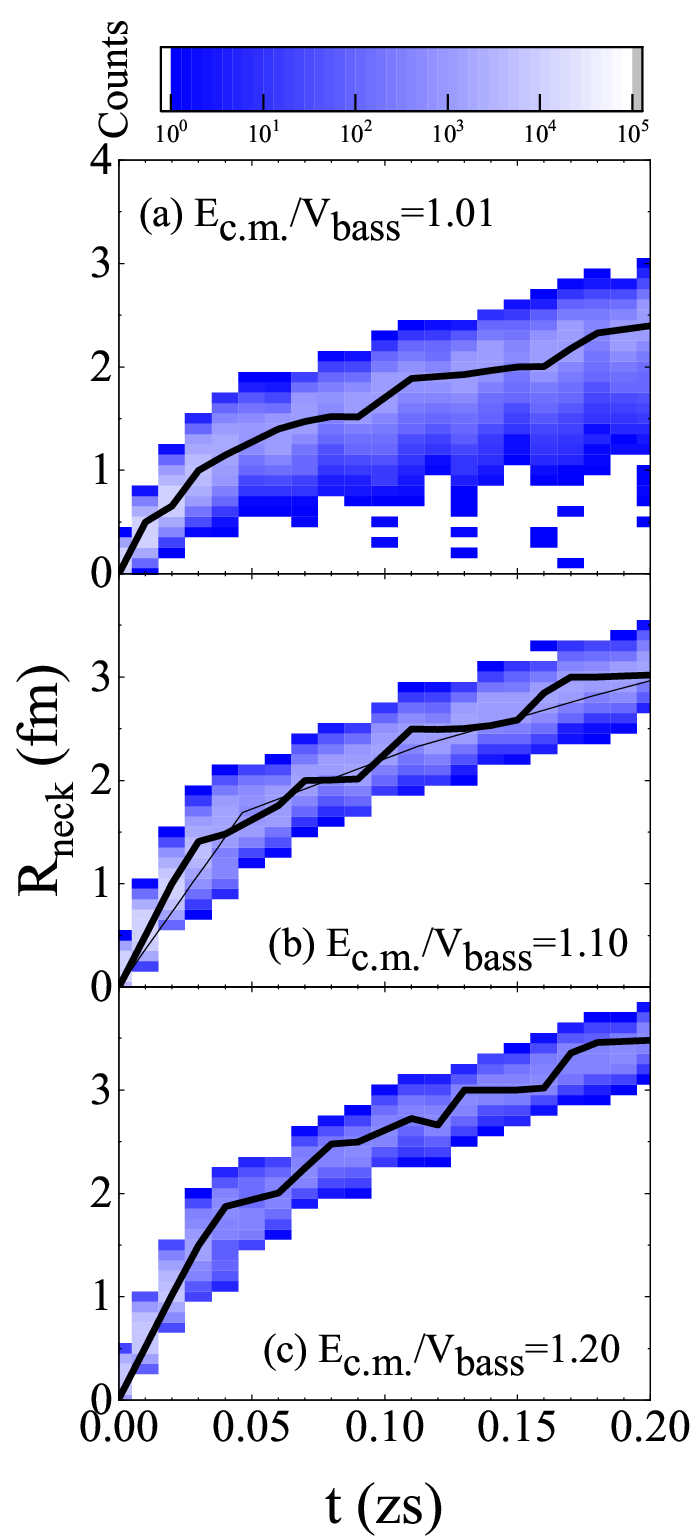}
\caption{Two-dimensional $R_\text{neck}$--time matrices for the $^{136}$Xe\,+\, $^{136}$Xe reaction at $E_\text{c.m.}/V_\text{bass}$ = (a) 1.01, (b) 1.10, and (c) 1.20. The black lines are the ridges of the contour map. The thin line is the calculation result of the neck radius with no fluctuation. The neck radii at each time step in 20000 calculations are accumulated to make contour maps.
}\label{fig4}
\end{figure}

\section{Results and discussion}\label{sec3}
\subsection{Neck growth in the early stage}

To analyze the dynamics of neck formation at the early stage of heavy-ion collisions, the fluctuation of the neck radius $R_\text{neck}$ versus evolution time for the $^{136}$Xe\,+\,$^{136}$Xe reaction is calculated. The calculations are performed at different centers of mass incident energies; $E_\text{c.m.}/V_\text{bass}$ = 1.01, 1.10, and 1.20 are calculated and are shown in Fig. 4, where $V_\text{bass}$ means the Bass barrier energy $V_\text{bass}$ = 299.56 $\text{MeV}$ \cite{bass1980nuclear}. The neck radius is extracted from each time step of the calculation and overplotted for 20000 trials. The values of neck radius are distributed as drawn by the contour map in Fig. 4, and the ridges are shown by the black lines. The presentations of calculations are restricted to the early stage of collision up to $0.2$ zs.
The rate of growth of the neck radius decreases beyond $t \approx$ 0.01, 0.03, and 0.04 zs for $E_\text{c.m.}/V_\text{bass}$ = 1.01, 1.10, and 1.20, respectively.
%Here, we defines the early stage up to $t \approx$ 0.01, 0.03 and 0.04 zs at $E_\text{c.m.}/V_\text{bass}=$ 1.01, 1.10 and 1.20 as the short-time region and at the stage longer than $t \approx$ 0.01, 0.03 and 0.04 zs as the long-time region.
Here, we define the region where the neck grows linearly as the viscous regime and the region thereafter as the inertial regime.

It is noted that in order to check the similarity with the neck formation of the macroscopic coalescence of water drop at first, the orbital angular momentum is set to zero in the nuclear collision system. When we estimate the viscosity in the later section, the finite vale of angular momentum corresponding to QF or DIC phenomena are used in the calculation.

%
%% amano added
In Fig. 4(a), the distribution spreads widely to the lower part because of the rapid disappearance of the neck owing to increasing components of DIC with decreasing incident energy.
The thin line in Fig. 4(b) represents the result with no fluctuation. Without fluctuation, the neck radius is uniquely determined.
As can be  seen in the black lines in Figs. 4(a)--(c), $R_\text{neck}$s grow linearly up to 0.01, 0.03, and 0.04 zs for each incident energy. This dependence on $E_\text{c.m.}$ is due to the extra kinetic energy of the projectile after contact.
The relative velocities of two colliding nuclei are 29, 31, and 32 fm/zs, corresponding to three incident energies. 
On the other hand, the growth rate of $R_\text{neck}$ decreases in the inertial regime.
The change in this slope signifies the transition of dynamics, as thoroughly discussed in  studies of water drops \cite{PhysRevLett.95.164503,PhysRevLett.98.224502,PhysRevLett.106.114501,doi:10.1073/pnas.1910711116,doi:10.1073/pnas.1017112108,AKELLA2020137917,thoroddsen_takehara_etoh_2005,article2}.
Here, the neck radius is estimated from the dam-bell shape of stuck nuclei.

We confirmed that these characteristics of neck formation in the very early stage can be seen even for the parameters $t_0$ and $\Delta_\epsilon$ of temporal function changing in the range 0.1 zs $\le t_0 \le 1$ zs and 0.1 zs $\le \Delta_\epsilon \le 10$ zs.

\subsection{Viscous-inertial crossover}\label{cross}
\begin{figure}[t]%fig5
\centering
\includegraphics[scale=0.53]{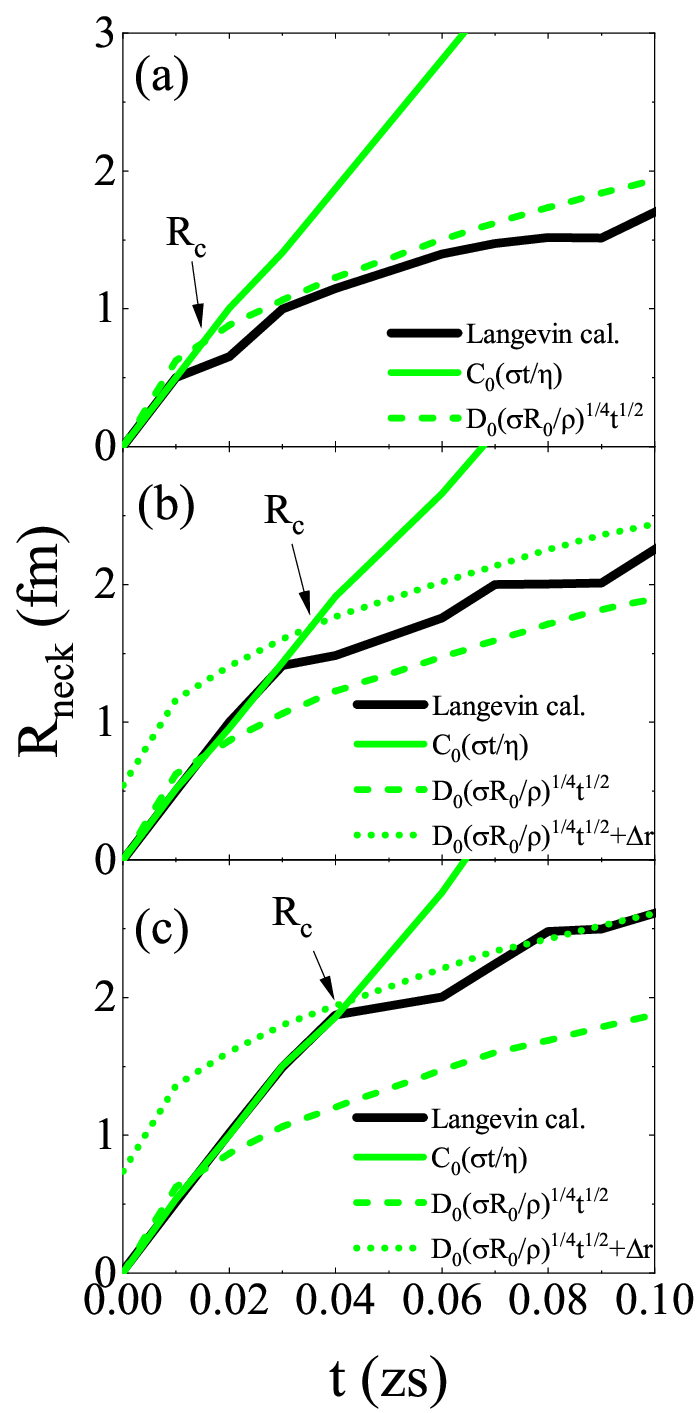}
\caption{Neck radius $R_\text{neck}$ during coalescence as a function of the time elapsed from the first connected neck for the $^{136}$Xe\,+\,$^{136}$Xe reaction at $E_\text{c.m.}/V_\text{bass}=$ (a) 1.01, (b) 1.10 and (c) 1.20. The solid black lines are the same as shown in Fig. 4. The green solid and dashed lines signify viscous-capillary ($R_\text{neck} \propto t$) and inertial-capillary scaling ($R_\text{neck} \propto t^{1/2}$). The dotted lines are the function of $R_\text{neck}$ considering the flow dynamics of the interacting region in nucleus--nucleus collisions. The correction factor $\Delta r$ and the coefficients $C_0$ and $D_0$ are explained in the text. The arrows indicate the crossover point $R_\text{c}$.
}\label{fig5}
\end{figure}
In the viscous regime, the capillary forces drive the growth of the neck against the viscosity. The function of the neck radius with respect to time $t$, $R_\text{neck}(t) \approx C_{0}\sigma t/\eta$ can be driven from the understanding that the decreased amount in the surface energy is dissipated by the viscosity \cite{PhysRevLett.106.114501, eggers_lister_stone_1999, doi:10.1073/pnas.1017112108}.
The value of the coefficient $C_{0}$ is 1.5 ($=3/2$) in this study.
%
%This idea also discussed in logarithmic corrections expressed $R_\text{neck}(t) \approx \gamma t/\pi \eta \ln (\gamma t/\eta R_\text{0})$ \cite{eggers_lister_stone_1999}.
%However, logarithmic corrections were not confirm as Fig. 2 in Ref. \cite{PhysRevLett.95.164503}.
Here, $\sigma$ shows the surface tension per unit area and the conventional value is used. $\eta$ denotes the viscosity of the nuclear matter.

On the other hand, when sufficient time elapses, the decrease in the surface energy is converted to the kinetic energy which corresponds to the beginning of the motion of the nucleons in the neck region.
Considering the balance of the kinetic energy with the surface energy, the neck radius is expressed as $R_\text{neck}(t) \approx D_{0}(\sigma R_{0}/\rho)^{1/4}\sqrt{t}$ \cite{duchemin_eggers_josserand_2003,eggers_lister_stone_1999} in the inertial regime.
%For the inviscid coalescence of drops, $R_\text{neck} \propto t^{1/2}$ is also observed in the numerical simulations \cite{duchemin_eggers_josserand_2003}.
$R_{0}=1.2A^{1/3}$ represents the nuclear radius with mass number $A$  before the collisions. $\rho$ is the nucleus density and is assumed to be $2.5 \times 10^{17}$ kg/m$^{3}$.
The value of the coefficient $D_{0}$ is approximately 1.19 ($\sim2^{1/4}$) in our study.
The values of both $C_{0}$ and $D_{0}$ are on the order of 1. These values are also reasonable for the analysis in our results, as reported in Refs. \cite{PhysRevLett.98.224502,PhysRevLett.106.114501, duchemin_eggers_josserand_2003}.
%

%%%%%%%
These two functions for neck are plotted in Fig. 5 with the calculation results indicated by the black lines in Fig. 4.
We resize the coordinate scales in Fig. 5 to focus on the transition of the dynamics.
From the data of linear dependence on time in the viscous regime in Fig. 5, we can extract the viscosity coefficient. The linear dependence of neck radius on time is drawn by the green solid line using the optimum value of $\eta$. The values of the viscosity coefficient extracted from Figs. 5(a), 5(b), and 5(c) are 0.0092 TP, 0.0090 TP, and 0.0086 TP, respectively.
%In our study, the estimated values are smaller than the viscosity of nucleus determined from the experimental data in GDR and fission processes.

%The linear growth of the neck by the viscous-capillary force was unclear in Fig. 5(a) but
%unlike in the case of the macroscopic drops. In contrast, the linear dependence of neck growth
%it becomes long lasting when the incident energy increases.
In the collisions of superheavy nuclei, an inner barrier exists even after overcoming the Coulomb barrier.
In the case of high-incident-energy reactions as in Figs. 5(b) and 5(c),
because of the incompressibility of nuclear matter, the compressive area thought to arise owing to the interaction with the inner barrier is evaded. Instead, the neck radius is considered to enlarge without changing the neck width. This effect is taken into account by the additional parameter $\Delta r$.
%where is the enlarged radius discussed later.
%{\bf the compressive area increases with incident energy in the moment of coalescence due to the interaction with the inner barrier. During this moment, because of the incompressibility of nuclei the neck region is considered to enlarge by about double size of overlap region without changing the neck width.}
In the experiment of the coalescence of droplets \cite{PhysRevLett.95.164503,PhysRevLett.98.224502,PhysRevLett.106.114501,doi:10.1073/pnas.1910711116,doi:10.1073/pnas.1017112108,thoroddsen_takehara_etoh_2005,article2}, the relative velocity between them is considered to be zero. In contrast, in our reaction system, the relative velocity of colliding nuclei is inevitable to overcome the Coulomb barrier for the coalescence.
In this point, for the present case the dominant motion in the neck region is different from the macroscopic case \cite{duchemin_eggers_josserand_2003} in the viscous regime.
%it is possible to approach an inner barrier with the contribution of the extra kinetic energy in the entrance channel after overcoming the coulomb barrier.
%Thus, the neck radius grows more, being able to coalesce.
%
%Therefore, under experimental conditions where the dominant motion is likely to differ between macroscopic and microscopic cases, the latter dynamics can be universally explained for the former dynamics by $\Delta r$.
This is the reason why the parameter $\Delta r$ is introduced above.

The crossover point $R_{c}$ has energy dependence, and the values are 0.6 fm, 1.6 fm, and 1.87 fm in Figs. 5(a)--(c), respectively.
From the corresponding shift of $R_\text{neck}(t)$ in the inertial regime, the values of  $\Delta r$ for the case of Figs. 5(b) and 5(c) are 0.54 and 0.74 fm.

\subsection{Viscosity coefficient for the case with large impact parameter}

%%%%%%%%%%%%%%%%%%%%%%%%%%%%%%%%%%%%%%%%%%%
\begin{figure}[t]%fig6
\centering
\includegraphics[scale=0.65]{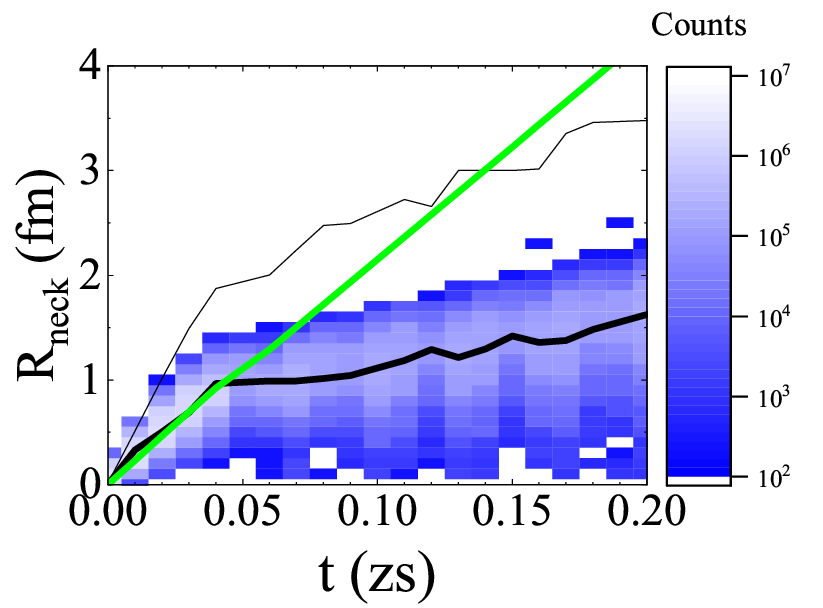}
\caption{Two-dimensional $R_\text{neck}$--time matrices for $L=210\hbar$ in the reaction of $^{136}$Xe\,+\, $^{136}$Xe at $E_\text{c.m.}/V_\text{bass}$ = 1.20. The thick black line is the ridges of the contour map. The neck radii at each time step in 900000 calculations are accumulated to make contour maps. The thin black line is the same as the result shown in Fig. 4(c). The green line signifies viscous-capillary ($R_\text{neck} \propto t$) scaling using $R_\text{neck}(t) \approx C_{0}\sigma t/\eta$.
}\label{fig6}
\end{figure}
%%%%%%%%%%%%%%%%%%%%%%%%%%%%%%%%%%%%%%%%%%%

We have carried out a new method to estimate the nuclear viscosity coefficient from the  neck formation.
In order to compare the viscosity coefficients from experimental data in the GDR, the calculations are performed for high orbital angular momentum ($L=210\hbar$). Figure 6 shows two-dimensional $R_\text{neck}$--time matrices for $L=210\hbar$ in the reaction of $^{136}$Xe\,+\, $^{136}$Xe at $E_\text{c.m.}/V_\text{bass}$ = 1.20. The distribution spreads widely to the lower part because of the rapid disappearance of the neck owing to increasing component of DIC comparing the result in Fig. 4(c).
The thin black line is the same as the black line in Fig. 4(c) calculated with $L=0\hbar$. At high orbital angular momentum, the rapid growth of the neck appearing in the viscous regime is suppressed and the slope is gentle due to less effect from the compressive region.
The change in neck radius is represented by the green line using the equation: $R_\text{neck}(t) \approx C_{0}\sigma t/\eta$.
When we fit the green line to the slope from 0 zs to 0.04 zs indicated by the thick black line, we obtain $\eta=0.02\pm0.01$ TP.
Next, we evaluate the value of the viscosity coefficient estimated using the friction tensor $\gamma_{\alpha\alpha}$ and the inertia tensor $m_{\alpha\alpha}$ treated in our model. The kinematic viscosity coefficient $\nu$ is shown using reduced friction coefficient $\beta_{\alpha\alpha}(q)$ as follows \cite{kar02,hern81,adeev05}

\begin{eqnarray}
&&\beta_{\alpha\alpha}\left(q\right)=\frac{\gamma_{\alpha\alpha}}{m_{\alpha\alpha}}, \\
&&\nu=\frac{\beta_{\alpha\alpha}\left(q\right)R_{neck}^2}{6},
\end{eqnarray}
$\eta$ treated as a two-body viscosity coefficient in nuclear physics is obtained using kinematic viscosity coefficient as $\eta=\rho\nu$. Averaging the values of $\eta$ between 0 zs and 0.04 zs where the viscous-capillary force is dominant, $\eta\approx0.021$ TP is obtained in the present case. The viscosity coefficient obtained from the reduced friction coefficient is almost consistent with the one obtained from the slope of the neck formation in our Langevin calculation.

\section{Conclusion}\label{sec4}

The dynamics of the growing neck radius in nucleus--nucleus collisions has been investigated by the unified Langevin equation method assuming the macroscopic mass parameter. We have chosen the $^{136}$Xe\,+\,$^{136}$Xe system as an example and focused on the early stage of collision, where the nuclear viscosity clearly controls the neck formation.
The values of the viscosity coefficient of the nucleus extracted in this study have been found to be $\eta=0.02\pm0.01$ TP, which is comparable to the viscosity determined from experimental data in GDR.
It is noted that the change in the dynamics occurred in the coalescence of nuclei is similar to the case of water drops. The viscous-capillary force ($R_\text{neck} \propto t$) works at short times, and the inertial-capillary force ($R_\text{neck} \propto t^{1/2}$) is dominant at later stage. It is remarkable that even in microscopic drops such as nuclei, the viscous-inertial crossover can be seen in the present simulation independent of the temporal form of neck parameter $\epsilon$.
%In the present simulation assuming the macroscopic mass parameter, the neck formation proceeds within $\sim$ 0.2zs and gives comparable value of $\eta$ to the estimated value from GDR phenomena.

%
%That is, the liquid drop model used in nuclear physics has an appropriate validity in the analysis of collective motions.
%
\section*{Acknowledgments}

The Langevin calculation was performed using the cluster computer system (Kindai-VOSTOK) under the supported of JSPS KAKENHI Grant Number 20K04003 and Research funds for External Fund Introduction 2021 provided by Kindai University.


\begin{thebibliography}{0}
%1
\bibitem{Thomson86} T. J. John and N. H. Frank, \href{https://royalsocietypublishing.org/doi/10.1098/rspl.1885.0034}{\color{blue}Proc. R. Soc. Lond {\bf39}, 239 (1886).}
%2
\bibitem{Weizs35} C. F. v. Weizs{\"a}cker, \href{https://doi.org/10.1007/BF01337700}{\color{blue}Z. Physik {\bf96}, 431 (1935).}
%3
\bibitem{Bethe36} H. A. Bethe and R. F. Bacher, \href{https://doi.org/10.1103/RevModPhys.8.82}{\color{blue}Rev. Mod. Phys. {\bf8}, 82 (1936).}
%4
\bibitem{PhysRev.56.426} N. Bohr and J. A. Wheeler, \href{https://link.aps.org/doi/10.1103/PhysRev.56.426}{\color{blue} Phys. Rev. {\bf56}, 426 (1939).}
%5
\bibitem{PhysRevLett.95.164503} D. G. A. L. Aarts, H. N. W. Lekkerkerker, H. Guo, G. H. Wegdam, and D. Bonn, \href{https://link.aps.org/doi/10.1103/PhysRevLett.95.164503}{\color{blue} Phys. Rev. Lett. {\bf95}, 164503 (2005).}
%6
\bibitem{PhysRevLett.98.224502} J. C. Burton and P. Taborek, \href{https://link.aps.org/doi/10.1103/PhysRevLett.98.224502}{\color{blue} Phys. Rev. Lett. {\bf98}, 224502 (2007).}
%7
\bibitem{PhysRevLett.106.114501} J. D. Paulsen, J. C. Burton, and S. R. Nagel, \href{https://journals.aps.org/prl/abstract/10.1103/PhysRevLett.106.114501}{\color{blue} Phys. Rev. Lett. {\bf106}, 114501 (2011).}
%8
\bibitem{Adam00} G.G.Adamian, N.V.Antonenko, A.Diaz-Torres, W.Scheid, \href{https://doi.org/10.1016/S0375-9474(99)00852-0}{\color{blue} Nucl. Phys. A {\bf671}, 233-254 (2000).}
%9
\bibitem{PhysRevLett.69.249} G. Enders, F. D. Berg, K. Hagel, W. K{\"u}hn, V. Metag, R. Novotny, M. Pfeiffer, O. Schwalb, R. J. Charity, A. Gobbi, R. Freifelder, W. Henning, K. D. Hildenbrand, R. Holzmann, R. S. Mayer, R. S. Simon, J. P. Wessels, G. Casini, A. Olmi, and A. A. Stefanini, \href{https://link.aps.org/doi/10.1103/PhysRevLett.69.249}{\color{blue} Phys. Rev. Lett. {\bf69}, 249 (1992).}
%10
\bibitem{BAUMANN1998428} T. Baumann, E. Ramakrishnan, A. Azhari, J. Beene,
R. Charity, J. Dempsey, M. Halbert, P.-F. Hua, R. Kryger, P. Mueller, R. Pfaﬀ, D. Sarantites, L. Sobotka, D. Stracener, M. Thoennessen, G. Van Buren, R. Varner, and S. Yokoyama, \href{https://doi.org/10.1016/S0375-9474(98)00197-3}{\color{blue} Nuclear Physics A {\bf635}, 428 (1998).}
%11
\bibitem{PhysRevLett.82.3404} M. P. Kelly, K. A. Snover, J. P. S. van Schagen,
M. Kici{\'n}ska-Habior, and Z. Trznadel, \href{https://link.aps.org/doi/10.1103/PhysRevLett.82.3404}{\color{blue} Phys. Rev. Lett. {\bf82}, 3404 (1999).}
%12
\bibitem{article} P. Heckman, D. Bazin, J. Beene, Y. Blumenfeld, M. Chromik, M. Halbert, F. Liang, E. Mohrmann, T. Nakamura, A. Navin, B. Sherrill, K. Snover, \href{https://doi.org/10.1016/S0370-2693(03)00017-0}{\color{blue} Physics Letters B {\bf555}, 43 (2003).}
%13
\bibitem{PhysRevLett.103.172501} N. Auerbach and S. Shlomo, \href{https://link.aps.org/doi/10.1103/PhysRevLett.103.172501}{\color{blue} Phys. Rev. Lett. {\bf103}, 172501 (2009).}
%14
\bibitem{PhysRevLett.70.3538} T. Wada, Y. Abe, and N. Carjan, \href{https://link.aps.org/doi/10.1103/PhysRevLett.70.3538}{\color{blue} Phys. Rev. Lett. {\bf 70}, 3538 (1993).}
%15
\bibitem{KOONIN197787} S. Koonin, R. Hatch, and J. Randrup, \href{https://www.sciencedirect.com/science/article/pii/0375947477907011}{\color{blue} Nuclear Physics A {\bf 283}, 87 (1977).}
%16
\bibitem{PhysRevC.17.646} A. J. Sierk, S. E. Koonin, and J. R. Nix, \href{https://link.aps.org/doi/10.1103/PhysRevC.17.646}{\color{blue} , Phys. Rev. C {\bf17}, 646 (1978).}
%17
\bibitem{PhysRevC.13.2385} K. T. R. Davies, A. J. Sierk, and J. R. Nix, \href{https://link.aps.org/doi/10.1103/PhysRevC.13.2385}{\color{blue} Phys. Rev. C {\bf 13}, 2385 (1976).}
%18
\bibitem{Zagrebaev_2005} V. Zagrebaev and W. Greiner, \href{https://doi.org/10.1088/0954-3899/31/7/024}{\color{blue} Journal of Physics G: Nuclear and Particle Physics {\bf 31}, 825 (2005).}
%19
\bibitem{PhysRevC.106.024610} S. Amano, Y. Aritomo, and M. Ohta, \href{https://link.aps.org/doi/10.1103/PhysRevC.106.024610}{\color{blue} Phys. Rev. C {\bf106}, 024610 (2022).}
%20
\bibitem{Zagrebaev_2007} V. Zagrebaev and W. Greiner, \href{https://doi.org/10.1088/0954-3899/34/11/004}{\color{blue} Journal of Physics G: Nuclear and Particle Physics {\bf 34}, 2265 (2007).}
%21
\bibitem{ARITOMO20043} Y. Aritomo and M. Ohta, \href{https://www.sciencedirect.com/science/article/pii/S0375947404008668}{\color{blue} Nuclear Physics A {\bf 744}, 3 (2004).}
%22
\bibitem{zagrebaev2007potential} V. Zagrebaev, A. Karpov, Y. Aritomo, M. Naumenko, and W. Greiner, \href{https://doi.org/10.1134/S106377960704003X}{\color{blue} Physics of Particles and Nuclei {\bf38}, 469 (2007).}
%23
\bibitem{bertsch1978collision} G. Bertsch, \href{https://doi.org/10.1007/BF01408501}{\color{blue} Zeitschrift f{\"u}r Physik A Atoms and Nuclei {\bf 289}, 103 (1978).}
%24
\bibitem{CASSING1983467} W. Cassing and W. N{\"o}renberg, \href{https://www.sciencedirect.com/science/article/pii/0375947483903615}{\color{blue} Nuclear Physics A {\bf 401}, 467 (1983).}
%25
\bibitem{PhysRevC.69.021603} A. Diaz-Torres, \href{https://link.aps.org/doi/10.1103/PhysRevC.69.021603}{\color{blue} Phys. Rev. C {\bf 69}, 021603(R) (2004).}
%26
\bibitem{maruhn1972asymmetrie} J. Maruhn and W. Greiner, \href{https://doi.org/10.1007/BF01391737}{\color{blue} Zeitschrift f{\"u}r Physik {\bf 251}, 431 (1972).}
%27
\bibitem{sato1978microscopic} K. Sato \textit{et al.}, \href{https://doi.org/10.1007/BF01417722}{\color{blue} Zeitschrift f{\"u}r Physik A Atoms and Nuclei {\bf 288}, 383 (1978).}
%28
\bibitem{YAMAJI1987487} S. Yamaji, H. Hofmann, and R. Samhammer, \href{https://www.sciencedirect.com/science/article/pii/0375947487900753}{\color{blue} Nuclear Physics A {\bf 475}, 487 (1987).}
%29
\bibitem{sai22} V. Saiko and A. Karpov, \href{https://rdcu.be/c8y2I}{\color{blue} Eur. Phys. J. A (2022) {\bf 58}, 41 (2022).}
%30
\bibitem{sai19} V. V. Saiko and A. V. Karpov, \href{https://link.aps.org/doi/10.1103/PhysRevC.99.014613}{\color{blue} Phys. Rev. C {\bf 99}, 014613 (2019).}
%31
\bibitem{kar17} A. V. Karpov and V. V. Saiko, \href{https://link.aps.org/doi/10.1103/PhysRevC.96.024618}{\color{blue} Phys. Rev. C {\bf 96}, 024618 (2017).}
%32
\bibitem{PhysRevC.20.992} H. J. Krappe, J. R. Nix, and A. J. Sierk, \href{https://link.aps.org/doi/10.1103/PhysRevC.20.992}{\color{blue} Phys. Rev. C {\bf 20}, 992 (1979).}
%33
\bibitem{STRUTINSKY19681} V. Strutinsky, \href{https://www.sciencedirect.com/science/article/pii/0375947468906994}{\color{blue} Nuclear Physics A {\bf 122}, 1 (1968).}
%34
\bibitem{RevModPhys.44.320} M. Brack \textit{et al.}, \href{https://link.aps.org/doi/10.1103/RevModPhys.44.320}{\color{blue} Rev. Mod. Phys. {\bf 44}, 320 (1972).}
%35
\bibitem{suek74} S. Suekane, A. Iwamoto, S. Yamaji, and K. Harada,
JAERI-memo, 5918 (1993).
%36
\bibitem{10.1143/PTP.55.115} A. Iwamoto \textit{et al.}, \href{https://doi.org/10.1143/PTP.55.115}{\color{blue} Progress of Theoretical Physics {\bf 55}, 115 (1976).}
%37
\bibitem{NILSSON19691} S. G. Nilsson \textit{et al.}, \href{https://www.sciencedirect.com/science/article/pii/0375947469908094}{\color{blue} Nuclear Physics A {\bf 131}, 1 (1969).}
%38
\bibitem{PhysRevC.90.054609} Y. Aritomo, S. Chiba, and F. Ivanyuk, \href{https://link.aps.org/doi/10.1103/PhysRevC.90.054609}{\color{blue} Phys. Rev. C {\bf 90}, 054609 (2014).}
%39
\bibitem{ignatyuk1975phenomenological} A. Ignatyuk, G. Smirenkin, and A. Tishin, \href{http://inis.iaea.org/search/search.aspx?orig_q=RN:06208426}{\color{blue} Yadernaya Fizika {\bf 21}, 485 (1975).}
%40
\bibitem{PhysRevC.80.064604} Y. Aritomo, \href{https://link.aps.org/doi/10.1103/PhysRevC.80.064604}{\color{blue} Phys. Rev. C {\bf 80}, 064604 (2009).}
%41
\bibitem{FROBRICH1998131} P. Fr{\"o}brich and I. I. Gontchar, \href{https://www.sciencedirect.com/science/article/pii/S0370157397000422}{\color{blue} Physics Reports {\bf 292}, 131 (1998).}
%42
\bibitem{BLOCKI1978330} J. Blocki \textit{et al.}, \href{https://www.sciencedirect.com/science/article/pii/0003491678902087}{\color{blue} Annals of Physics {\bf 113}, 330 (1978).}
%43
\bibitem{RAYFORDNIX1984161} J. R. Nix and A. J. Sierk, \href{https://www.sciencedirect.com/science/article/pii/0375947484902495}{\color{blue} Nuclear Physics A {\bf 428}, 161 (1984).}
%44
\bibitem{RANDRUP1984105} J. Randrup and W. Swiatecki, \href{https://www.sciencedirect.com/science/article/pii/0375947484901519}{\color{blue} Nuclear Physics A {\bf 429}, 105 (1984).}
%45
\bibitem{Feldmeier_1987} H. Feldmeier, \href{https://doi.org/10.1088/0034-4885/50/8/001}{\color{blue} Reports on Progress in Physics {\bf 50}, 915 (1987).}
%46
\bibitem{CARJAN1986381} N. C{\^a}rjan, A. J. Sierk, and J. R. Nix, \href{https://www.sciencedirect.com/science/article/pii/0375947486902046}{\color{blue} Nuclear Physics A {\bf 452}, 381 (1986).}
%47
\bibitem{carj92} N. C{\^a}rjan, T. Wada, and Y. Abe, AIP Conference Proceedings (1992).
%48
\bibitem{20067} T. Asano \textit{et al.}, \href{https://doi.org/10.14494/jnrs2000.7.7}{\color{blue} Journal of Nuclear and Radiochemical Sciences {\bf 7}, 7 (2006).}
%49
\bibitem{PhysRevC.21.982} A. J. Sierk and J. R. Nix, \href{https://link.aps.org/doi/10.1103/PhysRevC.21.982}{\color{blue} Phys. Rev. C {\bf 21}, 982 (1980).}
%50
\bibitem{bass1980nuclear} R. Bass, Nuclear reactions with heavy ions (1980).
%51
\bibitem{doi:10.1073/pnas.1910711116} X. Xia, C. He, and P. Zhang, \href{https://www.pnas.org/doi/abs/10.1073/pnas.1910711116}{\color{blue} Proceedings of the National Academy of Sciences {\bf116}, 23467 (2019).}
%52
\bibitem{doi:10.1073/pnas.1017112108} M. Yokota and K. Okumura, \href{https://doi.org/10.1073/pnas.1017112108}{\color{blue} Proceedings of the
National Academy of Sciences {\bf108}, 6395 (2011).}
%53
\bibitem{AKELLA2020137917} V. Akella and H. Gidituri, \href{https://doi.org/10.1016/j.cplett.2020.137917}{\color{blue} Chemical Physics Letters {\bf758}, 137917 (2020).}
%54
\bibitem{thoroddsen_takehara_etoh_2005} S. T. THORODDSEN, K. TAKEHARA, and T. G.
ETOH, \href{https://www.cambridge.org/core/journals/journal-of-fluid-mechanics/article/coalescence-speed-of-a-pendent-and-a-sessile-drop/21B93C2125304C05D92FADED6A8C0851}{\color{blue} Journal of Fluid Mechanics {\bf527}, 85–114 (2005).}
%55
\bibitem{article2} J. Paulsen, R. Carmigniani, A. Kannan, J. Burton, and S. Nagel, \href{https://www.nature.com/articles/ncomms4182}{\color{blue} Nature communications {\bf5}, 3182 (2014).}
%56
\bibitem{duchemin_eggers_josserand_2003} L. DUCHEMIN, J. EGGERS, and C. JOSSERAND, \href{https://www.cambridge.org/core/journals/journal-of-fluid-mechanics/article/inviscid-coalescence-of-drops/CB56EE5EDDAFF2E9503CFEB0D2110662}{\color{blue} Journal of Fluid Mechanics {\bf487}, 167–178 (2003).}
%57
\bibitem{eggers_lister_stone_1999} J. EGGERS, J. R. LISTER, and H. A. STONE, \href{https://www.cambridge.org/core/journals/journal-of-fluid-mechanics/article/coalescence-of-liquid-drops/0DE6817B486607109F5A32018951B5C0}{\color{blue} Journal of Fluid Mechanics {\bf401}, 293–310 (1999).}
%58
\bibitem{kar02} A. V. Karpov and G. D. Adeev, \href{https://doi.org/10.1134/1.1508691}{\color{blue} Phys. At. Nucl. {\bf 65}, 1596 (2002).}
%59
\bibitem{hern81} E. S. Hernandez, W. D. Myers, J. Randrup, and B. Remaud, \href{https://www.sciencedirect.com/science/article/abs/pii/0375947481906485}{\color{blue} Nucl. Phys. A {\bf 361}, 483 (1981).}
%60
\bibitem{adeev05} G. D. Adeev, A. V. Karpov, P. N. Nadtochii, and D. V. Vanin, \href{https://www.researchgate.net/publication/283364412_Multidimensional_stochastic_approach_to_fission_dynamic_of_excited_nuclei}{\color{blue} Physics of Particles and Nuclei {\bf 36(4)} {\bf401}, 378–426 (2005).}
\end{thebibliography}
\end{document}